\documentclass[twocolumn,aps,superscriptaddress,showpacs]{revtex4}
\usepackage{amssymb}
\usepackage{amsmath}
\usepackage{graphicx}
\usepackage[normalem]{ulem}
\usepackage[dvips]{color}
\usepackage{epsfig}

\newcommand{\be}{\begin{eqnarray}}
\newcommand{\ee}{\end{eqnarray}}

\def\lsim{\mathrel{\rlap{\lower3pt\hbox{\hskip1pt$\sim$}}
     \raise1pt\hbox{$<$}}} 
\def\gsim{\mathrel{\rlap{\lower3pt\hbox{\hskip1pt$\sim$}}
     \raise1pt\hbox{$>$}}} 

\def\la{\langle}\def\ra{\rangle}

\def\bi{\bibitem}

\begin{document}
\title{Holographic Vector Dominance for the Nucleon}
\author{Masayasu Harada}
\affiliation{Department of Physics,  Nagoya University,  Nagoya,
  464-8602, Japan}
\author{Mannque Rho}
\affiliation{Institut de Physique Th\'eorique,  CEA Saclay, 91191
  Gif-sur-Yvette C\'edex, France \&\\
Department of Physics, Hanyang University, Seoul 133-791, Korea}

\date{\today}

\begin{abstract}
We derive a two-parameter formula for the electromagnetic form factors
of the nucleon described as an instanton by ``integrating out" all KK
modes other than the lowest mesons from the infinite-tower of vector
mesons in holographic QCD while preserving hidden local symmetry for
the resultant vector fields. With only two parameters, the proton
Sachs form factors can be fit surprisingly well to the available
experimental data for momentum transfers $Q^2\lsim 0.5$ GeV$^2$ with
$\chi^2$/dof $\lsim 2$, indicating the importance of an infinite tower
in the soliton structure of the nucleon. The prediction of the
Sakai-Sugimoto holographic dual model is checked against the fit
values that we
interpret as representing Nature. The nature of the ``core" in the
nucleon structure is reassessed in light of the results obtained with
the infinite-tower structure.

\end{abstract}
\maketitle


The celebrated Sakurai vector dominance (sVD for short) model for the
EM form factors~\cite{sakurai} works fairly well for
mesons~\cite{sakurai-vd} at zero temperature and zero density but it famously fails for the
nucleon~\cite{IJL,BRW,bijker}. The failure has been interpreted as an
indication that the nucleon has a ``core" which is not present in the
pion structure~\cite{IJL,BRW}. The ``core" has been attributed -- among a
variety of sources -- to a compact microscopic structure of QCD
variables, such as for instance a little chiral bag with quarks
confined within~\cite{BRW}.

The recent development of holographic dual QCD (hQCD for short),
specially, the Sakai-Sugimoto string theory
model~\cite{sakai-sugimoto} that implements correctly chiral symmetry
of QCD, indicates a dramatic return of the notion of vector dominance
for {\em both} mesons and
baryons~\cite{HRYY,HRYY:VD,Hashimoto:2008zw,Kim:2008pw}. What
characterizes the baryon structure in the hQCD model is that the
baryon emerges as a soliton in the presence of an infinite tower of
vector mesons in addition to the pion. It is the infinite tower that
renders the VD description applicable to both mesons and baryons. In
fact, the isovector component of the EM form factor has the
``universal" form
\be
F_V^h=\sum_n^\infty \frac{g_{v^n}g_{v^nhh}}{Q^2+m_{v^n}^2}\label{FVhh}
\ee
where $g_{v^n}$ is the photon-$v^n$ coupling and $g_{v^nhh}$ is the
$v^n$-$hh$ coupling for $h=\pi, N$. Here $v^n$ is the $n$-th
isosvector vector mesons $\rho, \rho^\prime, ...$ in the tower. What
distinguishes one from the other is the vector-meson coupling to the
hadron. This form -- which turns out to be almost completely saturated
by the first four vector mesons -- works surprisingly well for both
mesons and baryons at low-momentum transfers, say, $Q^2\lsim 0.5$
GeV$^2$.

Three questions arise from these results.

The first is what makes the difference between the ``good" sVD for the
pion and the ``bad" sVD for the nucleon disappear when the infinite
tower is present?

The second is the problem of the ``core." While it is reasonable to
view the pion as point-like when probed at long wavelength, the
nucleon is an extended object, for which a local field description
must break down at some -- not too high -- momentum scale. There are
indications from high-energy proton-proton scattering, experiments on
high mass muon pairs and also in deep inelastic scattering off nucleon
that the nucleon has a core of $\sim 0.2 - 0.3$ fm in
size~\cite{core}. It is this class of observations that led to the
notion of the ``Little Bag"~\cite{BRlittlebag} and to the hybrid
structure of the nucleon with a core made up of a quark bag surrounded
by a meson cloud~\cite{BRW}. The question is whether the core is
subsumed in the infinite tower and if so, where and in what
form. Closely tied to this question is: Does the photon ``see" the
size of the instanton -- which is a skyrmion in the infinite tower of
vector mesons -- which goes as $\sim 1/\sqrt{\lambda}$ where
$\lambda=N_c g_{YM}^2$ is the 't Hooft constant ? For a large value of
$\lambda$, the soliton(instanton) size is small. However this size
cannot be a physical quantity. The physical size should be independent
of $\lambda$, which is akin to the bag size which is also unphysical
according to the Cheshire Cat principle~\cite{Kim:2008pw}. So the last
question is: Is the ``core" a physical observable?

The objective of this Letter is to address the above three questions.

We first define the notations that we shall use and then give a
concise summary of the hQCD calculation of the form factors as
described in \cite{HRYY,HRYY:VD,Hashimoto:2008zw}. We shall follow the
notations of \cite{HRYY,HRYY:VD}.

The nucleon form factors are defined from the matrix elements of the
external currents,
\be
\la p^{\prime}|J^{\mu}(x)|p\ra=e^{iqx}\,\bar
u(p^{\prime})\,{\cal O}^{\mu}(p,p^{\prime})\,u(p) \:,
\ee
where $q=p^{\prime}-p$ and $u(p)$ is the nucleon spinor of momentum
$p$.
From the Lorentz invariance and the current conservation, with the
assumption of CP invariance, the operator ${\cal O}^{\mu}$ takes the
form  (say, for the EM current)
\be
{\cal
O}^{\mu}(p,p^{\prime})
&=&
\gamma^{\mu}\left[\frac12F_1^s(Q^2)+F^v_1(Q^2)\tau^3\right]
\nonumber\\
&+&
i\frac{\sigma^{\mu\nu}}{2m_N}q_{\nu}
\left[\frac12F_2^s(Q^2)+F^v_2(Q^2)\tau^3\right]\:,
\ee
where $m_N$ is the nucleon mass and $\tau^a=\sigma^a/2$.
$F_1^{s,v}$ and $F_2^{s,v}$ are the Dirac and Pauli form factors
for isoscalar ($s$) current and isovector ($v$) respectively.

As matrix elements, the form factors contain all one-particle
irreducible diagrams
for two nucleons and one external current.  
Thus they are very difficult to calculate in QCD. It turns out,
however, that the anti-de
Sitter/conformal field theory (AdS/CFT) correspondence, or
gravity/gauge theory correspondence, found in
certain types of string theory, enables one to compute such
non-perturbative quantities as hadron form factors within certain
approximations.

According to the AdS/CFT correspondence, the low energy effective
action of the gravity
dual in the bulk sector of QCD becomes the generating functional for
the correlators in the gauge sector
of operator ${\cal O}$ in QCD in the large $N_c$ limit. The
normalizable modes of the bulk field are identified as the physical
states
in QCD, created by the operator ${\cal O}$.

The model we shall use is the gravity dual of low energy QCD  with
massless flavors in the large $N_c$ (or quenched) approximation
constructed by Sakai and Sugimoto (SS)~\cite{sakai-sugimoto}.
The holographic dual of spin-$1\over 2$ baryons, or nucleons, for two
flavors that we shall adopt in this paper was constructed in
\cite{HRYY} by introducing a bulk baryon field, whose effective action
is given in the ``conformal coordinate" $(x,w)$ as
\be
S_{5D}^{\rm eff}=&&\int_{x,w}(-i\bar{\cal B}\gamma^m D_m {\cal B}
-i m_b(w)\bar{\cal B}{\cal B} \nonumber\\
&+&\kappa(w)
\bar{\cal B}\gamma^{mn}F_{mn}^{SU(2)_I}{\cal B}+\cdots )+S_{\rm meson},
\label{5dfermion1}
\ee
where ${\cal B}$ is the 5D bulk baryon field, $D_m$ is the gauge
covariant derivative
and $S_{\rm meson}$ is the effective action for the mesons. Using the
instanton nature of baryon, the coefficients $m_b(w)$ and $\kappa(w)$
can be reliably calculated in string theory. In (\ref{5dfermion1}) the
ellipsis stands for higher derivative terms that
are expected to be suppressed at low energy, $E<M_{KK}$, where the KK
mass sets the cut-off scale. Note that the magnetic coupling involves
only the non-abelian part of the flavor symmetry $SU(2)_I$ -- with
abelian $U(1)_B$ being absent -- due to the non-abelian nature of
instanton-baryons.

The (nonnormalizable) photon field is written as
\begin{equation}
A_{\mu}(x,w)=\int_q\,A_{\mu}(q)A(q,w)\,e^{iqx}\:,
\end{equation}
with boundary conditions
that $A(q,w)=1$ and $\partial_wA(q,w)=0$ at the UV boundary, $w=\pm
w_{\rm max}$ and
the (normalizable) bulk baryon field as
\begin{equation}
{\cal B}(w,x)=\int_p\left[f_L(w)u_L(p)+f_R(w)u_R(p)\right]e^{ipx}\,.
\end{equation}
These 5D wave functions, $A(q,w)$ and $f_{L,R}(w)$, are determined by solving the equation of motion from our action (\ref{5dfermion1}).
Then, using the AdS/CFT correspondence, one can read off the Dirac form factor and the Pauli form factor~~\cite{HRYY,Hashimoto:2008zw},
\begin{eqnarray}
F_1^p(Q^2) = \sum_{k=0}^{\infty}
  \frac{g_V^{(k)} \, g_{v^k}}{Q^2 + m_{2k+1}^2},
F_2^p(Q^2) = \frac{1}{2}\,\sum_{k=0}^{\infty}
  \frac{g_2^{(k)} \, g_{v^k}} {Q^2 + m_{2k+1}^2}
\nonumber\\
\label{EM infinite}
\end{eqnarray}
with
\begin{eqnarray}
g_V^{(k)} =&&\int_{-w_{max}}^{w_{max}} dw\,\left|f_L(w)\right|^2\nonumber\\
&& \times (\psi_{(2k+1)}(w)+\kappa(w)  \partial_w \psi_{(2k+1)}(w))\label{gVk}\\
 g_2^{(k)}/4m_N &=&
  \int_{-w_{max}}^{w_{max}}
dw \,\kappa(w) f_L^*(w)f_R(w) \psi_{(2k+1)}(w)\label{g2k}
\end{eqnarray}
where $g_{v^k} = \zeta_k \, m_{2k+1}^2$
is the $v^k$ decay constant and $\psi_{(2k+1)}$ is the
wave function of the vector meson given in
\cite{sakai-sugimoto}. (Here $g_V^{(k)}$ is $g_{v^k NN}$ of
(\ref{FVhh})). This is the set of formulas that we shall use for our
analysis that follows.

For comparison with experiments, it is more convenient to work with
the Sachs form factors. They are given in terms of the Dirac and Pauli
form factors defined above as
\begin{eqnarray}
G_E^p(Q^2) &=& F_1^p(Q^2) - \frac{Q^2}{4 m_N^2} \, F_2^p(Q^2)
\label{GE}
\ ,
\\
G_M^p(Q^2) &=& F_1^p(Q^2) + F_2^p(Q^2).\label{GM}
\end{eqnarray}

In order to focus on the role that the lowest vector mesons
$V^{(0)}\equiv (\rho,\omega)$ play in the form factors and expose the
importance of the tower, we would like to integrate out all other
vector mesons than the $V^{(0)}$. In doing this, it is important to
preserve hidden local symmetry for $V^{(0)}$
as well as chiral symmetry.
How this can be done has
been worked out for the
pion~\cite{Harada:2006di,Harada:2010cn}.

  Following the strategy proposed in \cite{Harada:2010cn} for the pion
form factor, we first obtain an effective 4D Lagrangian for
an infinite tower of KK modes and the nucleon
from the
5D one in (\ref{5dfermion1}) by performing the integration over $w$.
Then, we integrate out all KK modes other than
$V^{(0)}$ through the equations of motion for the
higher KK modes.
This procedure gives an effective 4D Lagrangian to
${\mathcal O}(p^4)$
for
$V^{(0)}$ and the nucleon invariant under the hidden local
symmetry as well as chiral symmetry.
Going to higher orders cannot be
justified unless one incorporates higher order terms in the DBI action
and loop corrections in the bulk sector which is not at present
doable. What we are doing is essentially to ``integrate out" the tower
{\em at the tree level} with the effects of the higher tower lodged in
the action given to ${\cal O}(p^4)$. It is in this sense that we are
exposing the infinite tower effect as {\em corrections} to the lowest
KK mode. The validity of such a procedure is clearly limited to low
momentum transfer. We are thus limiting to $Q^2\lsim 0.5$ GeV$^2$.

An important point to note here is that the formula we derive is quite
generic for low energy processes cut off at the KK mass ($M_{KK}$)
scale (or equivalently at the chiral scale $\Lambda_\chi\sim 4\pi
f_\pi$). For the generic formula (\ref{GE form}) that is deduced from (\ref{FVhh}),
what matters is the infinite tower structure, which could be
``dimensionally deconstructed" bottom-up~\cite{SS} or reduced \`a la
Klein-Kaluza from string theory top-down.

We shall show explicitly in an extended version how the reasoning made
in \cite{Harada:2010cn} for the pion form factor can be taken over for
the nucleon case.  It is however not difficult to see intuitively what
happens. As shown in \cite{HRYY}, the crux of the matter is that the
direct photon coupling to the instanton can be eliminated by a field
redefinition of the photon field and the charge sum rule, so the form factor is entirely
vector dominated by the infinite tower just as the pion form factor
is. The only difference is in $g_{v^n hh}$, namely the coupling of the
vector meson $v^n$ to hadron $h=\pi, N$. The instantonic structure of
the baryon is then buried in the coupling $g_{v^n NN}$. It can be
shown that when all the KK modes other than the lowest one $V^{(0)}$
are integrated out, the resultant Sachs form factors take the form
\be
G_{E,M}^p (Q^2)/\beta_{E,M}
&=& \left( 1 - \frac{a_{E,M}}{2} \right) + z_{E,M}\,
  \frac{Q^2}{m_\rho^2}
\nonumber\\
 &+& \frac{a_{E,M}}{2} \, \frac{m_\rho^2}{m_\rho^2 + Q^2 }
\ ,
\label{GE form}
\ee
where $\beta_E=1$ and $\beta_M=\mu_p$, and  $a_{E,M}$ and $z_{E,M}$
are parameters to be determined.
It follows from arguments in complete parallel to those given in
\cite{Harada:2010cn} for the pion form factor that formally
integrating out higher modes is equivalent to expanding
(\ref{EM infinite}) in $Q^2/m_{2k+1}^2$ for $k \geq 1$ up to
${\mathcal O}(Q^2)$
while keeping the $\rho$ meson ($k=0$ mode) propagator as it is in
consistency with perturbative unitarity.

In what follows we do this expansion explicitly for the electric form
factor, but the same holds for the magnetic form factor.
The result for the electric form factor is
\begin{eqnarray}
G_E^p(Q^2)
&= & g_V^{(0)}\frac{m_{0}^2 }
   {Q^2 + m_{0}^2} +
  \left[ 1 - g_V^{(0)}\right]\nonumber\\
 {} &-& \frac{Q^2}{m_{0}^2}\,
  \sum_{k=1}^{\infty} g_V^{(k)}\frac{\zeta_k \, m_{0}^2 }{m_{2k+1}^2}
 - \frac{Q^2}{8 m_N^2} g_2 \label{GE-expand}
\end{eqnarray}
where we have used the sum rules
$\sum_{k=0}^{\infty} g_V^{(k)} \zeta_k= 1$ and
$\sum_{k=0}^{\infty}   g_2^{(k)} \, \zeta_k = g_2$.
Comparing (\ref{GE form}) and (\ref{GE-expand}), we find the $a$ and
$z$ parameters for the electric form factor
\begin{eqnarray}
a_E^{\rm (hQCD)} &=& 2 g_V^{(0)}\zeta_0
\ ,
\\
z_E^{\rm (hQCD)} &=& -
  \sum_{k=1}^{\infty} g_V^{(k)}\frac{\zeta_k \, m_{0}^2 }{m_{2k+1}^2}
  - \frac{m_{0}^2}{8 m_N^2} \, g_2 \ .
\label{z hQCD}
\end{eqnarray}

Making a similar expansion for the magnetic form factor, one finds
from (\ref{EM infinite})
\begin{eqnarray}
a_M^{\rm (hQCD)} &=&
 \left[ 2 g_V^{(0)}\zeta_0
  + g_2^{(0)} \zeta_0
 \right]
 / \mu_p
\ ,
\\
z_M^{\rm (hQCD)} &=& -
\sum_{k=1}^{\infty} \left[g_V^{(k)}
  + \frac{1}{2} g_2^{(k)}
\right]
\frac{m_{0}^2 \zeta_k}{m_{2k+1}^2}
/ \mu_p
\ ,
\label{z m hQCD}
\end{eqnarray}
with $\mu_p = 1 + (1/2) g_2$.

We now turn to the analysis of Eq.(\ref{GE form}) done in three
different ways. It is in the $a$ and $z$ parameters that hadron
structure figures.

First we consider the parameters $(a,z)$ as {\em totally free} and
best-fit (\ref{GE form}) to the accurate experiments given in
Ref.~\cite{Arrington:2007ux}. Given that the approximation is valid
for low momentum transfers, we limit to $Q^2\leq 0.5$
GeV$^2$. Throughout we use the values
\begin{equation}
m_\rho = 0.775 \,\mbox{GeV}
\ ,
m_N = 0.938 \, \mbox{GeV}
\ .
\end{equation}

\begin{figure}[htbp]
\begin{center}
\includegraphics[width=4.2cm]{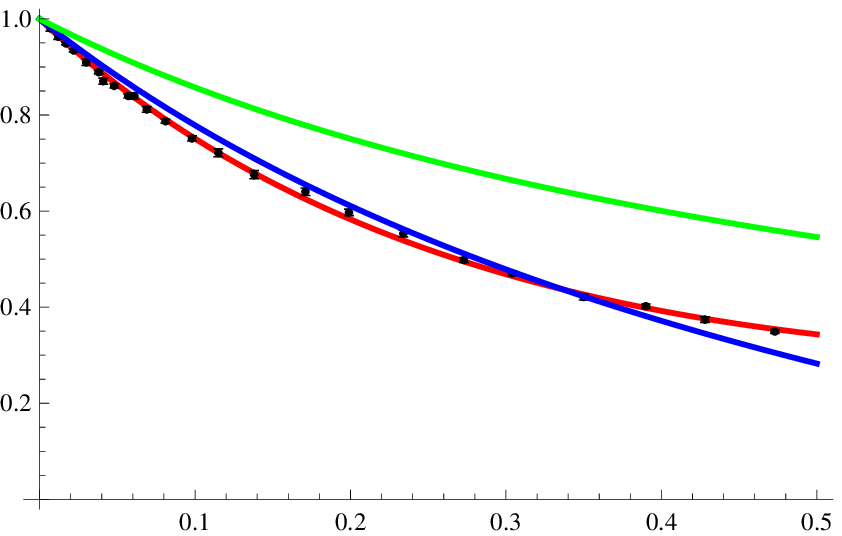}
\includegraphics[width=4.2cm]{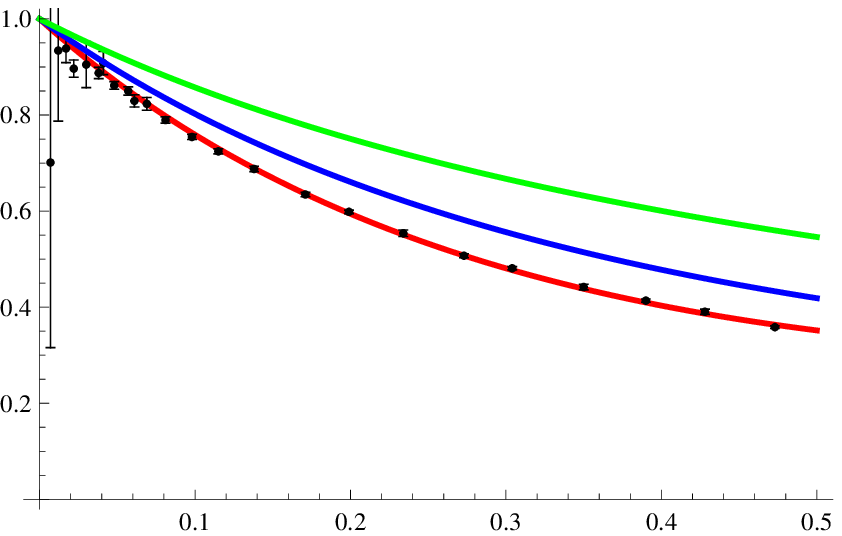}
\end{center}
\caption[]{(color online) $G_E^p$ (left panel) and $G_M^p/\mu_p$
  (right panel) vs. $Q^2$. The horizontal axis is $Q^2$ in
units of $\mbox{GeV}^2$. The red, green and blue curves are,
  respectively, the ``best fit," the sVD prediction and the hVD
  prediction.
}\label{fig gegd0.5}
\end{figure}

\begin{figure}[htbp]
\begin{center}
\includegraphics[width=4.2cm]{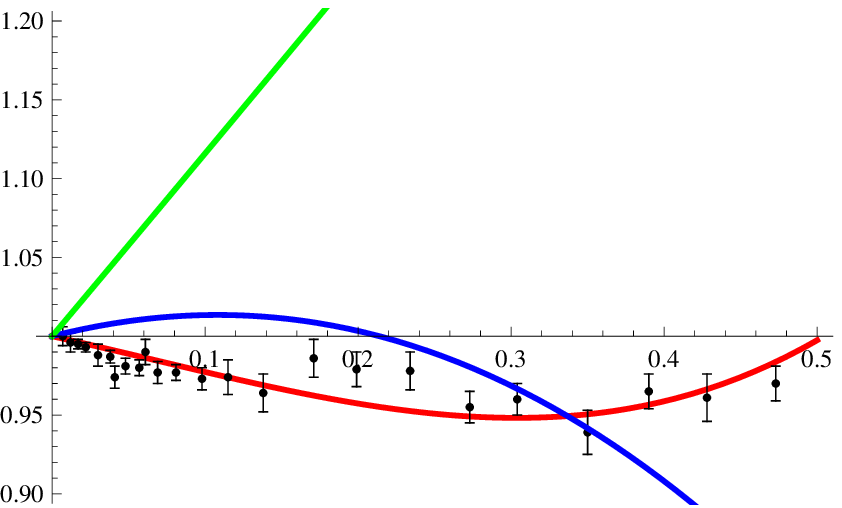}
\includegraphics[width=4.2cm]{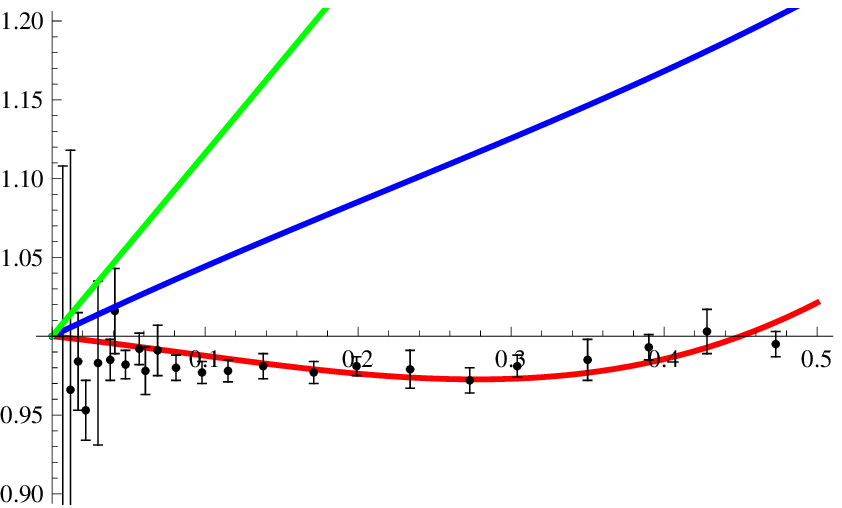}
\end{center}
\caption[]{(color online) $G_E^p/G_D$ (left panel) and
  $G_E^p/\mu_pG_D$ (right panel) vs. $Q^2$. See Fig.~\ref{fig gegd0.5}
  for labels.}\label{fig ge0.5}
\end{figure}

The best $\chi^2$ fit is given in red in Fig \ref{fig gegd0.5} for the
form factors and in Fig \ref{fig ge0.5} for the same divided by the
dipole form factor
$G_D(Q^2)=\left(1+\frac{Q^2}{0.71\ {\rm GeV}^2}\right)^{-2}$
which has been proven to be fairly close to
Nature.
The best fit parameters and $\chi^2$'s come out to be
\be
G_E^p:&&
a_E^{(best)}=4.55, z_E^{(best)}=0.45; \chi_E^2/{\rm dof}=1.5
\label{bestGE}\\
G_M^p:&&
a_M^{(best)}=4.31, z_M^{(best)}=0.40; \chi_M^2/{\rm dof}=1.1.
\label{bestGM}
\ee
Since the fit parameters are close to each other with similar
$\chi^2$, we may assume that $a_E=a_M$ and $z_E=z_M$ although we have
no compelling reason to do so, and make the best-fit. The result is:
$a^{(best)} =  4.42$ and $z^{(best)} =  0.42$ with $\chi^2$/dof =
1.90, quite close to (\ref{bestGE}) and (\ref{bestGM}). It is
interesting to compare the best-fit for the nucleon to the best-fit
for the pion form factor as obtained in \cite{Harada:2010cn}:
$a_\pi^{(best)}=2.44$ and $z_\pi^{(best)}=0.08$ with $\chi^2$/dof=2.44
while the sVD with $a=2$ and $z=0$ gives $\chi^2$/dof=4.3. Note that
in the case of the pion, the deviation in $\chi^2$ of the best-fit
from the sVD is relatively small accounting for the general acceptance
of the sVD.

We now examine how the vector dominance
models, sVD and hVD, fare in predicting these parameters and in
fitting the data for the nucleon.

First we consider the sVD. It has been known that the sVD does not
work at all for the nucleon although it has been fairly successful for
mesons. The sVD corresponds to taking in (\ref{GE form})
\be
a_E^{VD} &=&2, z_E^{VD}=-m_\rho^2/(4 m_N^2)\simeq -0.171,
\label{VDE}\\
a_M^{VD} &=&2, z_M^{VD}=0.\label{VDM}
\ee
(The second term in (\ref{VDE}) comes from the second term of
(\ref{GE})). The result is shown in green in Figs.~\ref{fig gegd0.5}
and \ref{fig ge0.5}. The $\chi^2$/dof comes out to be 187 and 852,
respectively, for $G_E^p$ and for $G_M^p$. This result merely confirms
the well-known story that sVD does not work for the nucleon.

Now turning to hVD, we will use the results of \cite{HRYY}.
We are limiting to the lowest four states since it is found
numerically that the charge and magnetic sum rules are almost
completely saturated by them~\cite{HRYY:VD}. One should however be
careful in using this observation for form factors since the four
states may not saturate momentum-dependent observables as fully as the
static quantities.

By using the values listed in Table~2 of the first reference of \cite{HRYY}, the parameter
$a$ comes out to be
\begin{eqnarray}
a_E^{\rm (hQCD)} = 3.01 \ ,
a_M^{\rm (hQCD)} = 3.14 \ ,
\label{a m v hQCD}
\end{eqnarray}
As for $z_{E,M}$, there are no known sum rules for the sums in
Eqs.~(\ref{z hQCD}) and (\ref{z m hQCD}). We shall simply take the
values for $k=1,2,3$ from the table
\begin{eqnarray}
z_E^{\rm (hQCD)} &\simeq& -
  \sum_{k=1}^{3} g_V^{(k)} \frac{\zeta_k \, m_{0}^2 }{m_{2k+1}^2}
  - \frac{m_{0}^2}{8 m_N^2} \, g_2 = -0.042 \ ,
\label{z v hQCD}
\\
\mu_p z_M^{\rm (hQCD)} &\simeq& -
\sum_{k=1}^{3} g_V^{(k)} \zeta_k
  + \frac{1}{2} g_2^{(k)} \zeta_k
\frac{m_{0}^2 }{m_{2k+1}^2} = 0.16
\ .
\label{z m v hQCD}
\end{eqnarray}

The form factors predicted in hQCD using the constants so determined,
i.e., $(a_E,z_E)=(3.01, -0.042), \ \ (a_M,z_M)=(3.14, 0.16)$,
are plotted in blue in Figs.~~\ref{fig gegd0.5} and \ref{fig
  ge0.5}. The $\chi^2$ comes out to be 20.2 for $G_E^p$ and 133 for
$G_M^p$. While $G_E$ comes out to be reasonable, $G_M$ is not.

To conclude, while the infinite tower structure of hQCD model gives
the pion form factor that deviates
little from the Sakurai VD, the nucleon form
factor deviates drastically from it. In the way the form factor is
parameterized, the $V^{(0)}=\rho,\omega$ contribution to the
nucleon form factor is considerably greater than that to the
pion, so the conventional VD with the ``universality" does not hold at
all. However this feature suggested by Nature is reproduced
semi-quantitatively in the hQCD model of Sakai and Sugimoto. For
instance, the charge contributed from the $V^{(0)}$ overshoots the proton charge by more than 40\%, which then is
dominantly compensated by the negative charge contribution coming from
the first excited vector mesons. This requires the so-called ``core"
contribution to be large but of the sign opposite to
what has been associated with a possible short-distance QCD
component~\cite{IJL,BRW,bijker}. This result brings us to question the {\em standard} interpretation of
the experimental observation~\cite{core} as an evidence for a microscopic QCD degree of freedom.

The deviation from the fit values of the parameters $(a,z)$  in the hQCD prediction for the nucleon, particularly in the magnetic form factor, may be pointing to the potential importance of what is manifestly missing in the hQCD model, namely, $1/N_c$ corrections, quark-mass effects and/or certain curvature effects in the bulk sector. This is an open issue in the field.

This work was partially supported by the WCU project of Korean
Ministry of
Education, Science and Technology (R33-2008-000-10087-0).
M.H. was supported in part by
Nagoya University Global COE Program,
the JSPS Grant-in-Aid for Scientific Research (S) \#22224003
\&
(c) \#20540262 and
Grant-in-Aid for Scientific Research on Innovative Areas (\#2104)
``Quest on New Hadrons with Variety of Flavors'' from MEXT.




\end{document}